\documentclass[11pt]{diazessay} % Font size (can be 10pt, 11pt or 12pt)

\title{\textbf{On-orbit No-contact Anomaly Debug Procedure for the CuPID Cubesat} \\ {\Large\itshape Root cause investigation}} % Title and subtitle

\author{\textbf{Emil A. Atz\\ Brian M. Walsh \\Connor J. O'Brien} \\ \textit{Boston University\\Center for Space Physics\\Mechanical Engineering}} % Author and institution

\date{\today} % Date, use \date{} for no date

\begin{document}

\maketitle % Print the title section

%----------------------------------------------------------------------------------------
%	ABSTRACT AND KEYWORDS
%----------------------------------------------------------------------------------------

%\renewcommand{\abstractname}{Summary} % Uncomment to change the name of the abstract to something else

\begin{abstract}
The CuPID CubeSat Observatory was a 6U cubesat launched into low-Earth orbit with a ride-share opportunity in Fall 2021. The mission was supported by NASA's Heliophysics division and motivated scientifically with the objective to image X-rays produced in the magnetosphere. After launch, the team was unable to communicate with the spacecraft. This document presents the testing and analysis of attempts to contact the spacecraft and investigation to the likely cause of failure with the radio system.
\end{abstract}

\hspace*{3.6mm}\textit{Keywords:} Cubesat, no-contact, anomaly, on-orbit test% Keywords

\vspace{30pt} % Vertical whitespace between the abstract and first section

%----------------------------------------------------------------------------------------
%	ESSAY BODY
%----------------------------------------------------------------------------------------

\section*{Introduction}
\subsection*{Mission summary}
The Cusp Plasma Imaging Detector (CuPID) CubeSat Observatory was developed by a multi-disciplinary team at Boston University, NASA Goddard Space Flight Center, Johns Hopkins University, Drexel University, Merrimack College, and the University of Alaska, Fairbanks. The project was supported through NASA's Heliophysics Technology and Instrument Development for Science (H-TIDeS) program. The mission was motivated scientifically to study temporal and spatial features of magnetic reconnection by imaging plasma density structures in soft X-rays in the magnetospheric cusps. CuPID was funded in 2016 for an estimated launch in 2020, though delay in launch manifest and the CoVID-19 pandemic delayed launch until 2021. 

The launch as a ride-share with the LandSat-9 mission was supported through NASA's CubeSat Launch Initiative (CSLI). The spacecraft was delivered to the launch vehicle in July 2021 and launched on September 27, 2021 into a nearly sun-synchronous orbit (97.6$^\circ$ inclination) at 550 km altitude.

\begin{figure*}
\includegraphics[width=300px]{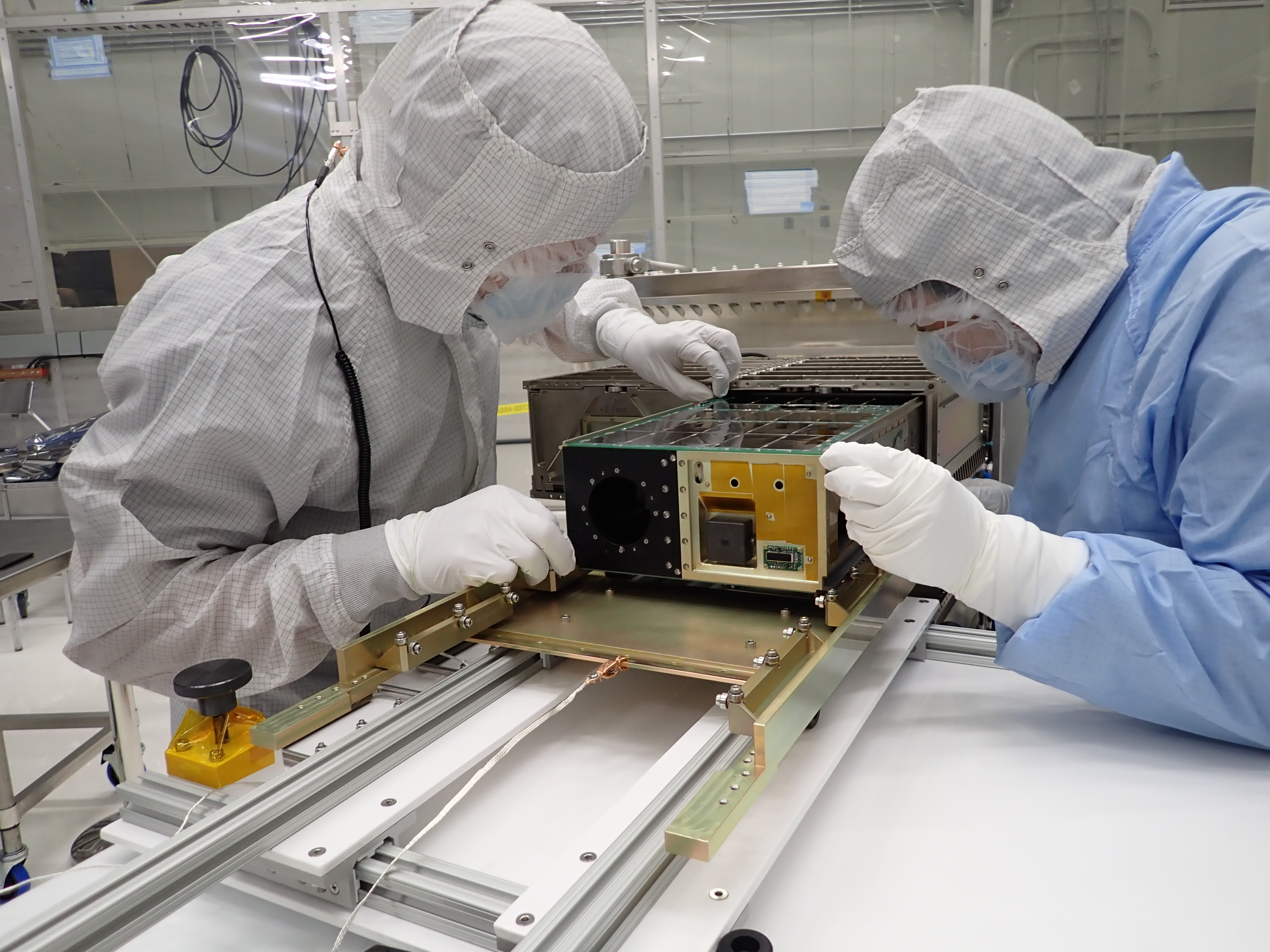}
\centering
\caption{\label{Integrate}Two students install CuPID into the ESPA (EELV Secondary Payload Adapter) ring CubeSat deployer during integration with launch vehicle components.}
\end{figure*}

\subsection*{Document overview}
With CubeSats of today being a system of systems, we first describe the CuPID mission and critical elements of the satellite important to the no-contact investigation. Then, the anomaly testing sections detail the possible issues inside these flight critical elements and the steps taken to test, and remedy them where possible.

\section*{CuPID Mission Design}
To meet its science objectives, the CuPID cubesat carried two co-aligned science instruments with fixed geometries. At high latitudes, the three-axis stabilized spacecraft adjusted its attitude to point the science instruments radially outward (zenith) and image the magnetospheric cusps. More about the mission design can be found in Walsh et al, 2021 \cite{walsh2021cusp}. The instruments included a wide field-of-view X-ray telescope, micro-dosimeter suite and body mounted magnetometer utilized for attitude determination and magnetic field monitoring. More details on the instruments and calibrations are found in Atz et al, 2022 \cite{atz2022cusp}.

\begin{figure*}
\includegraphics[width=350px]{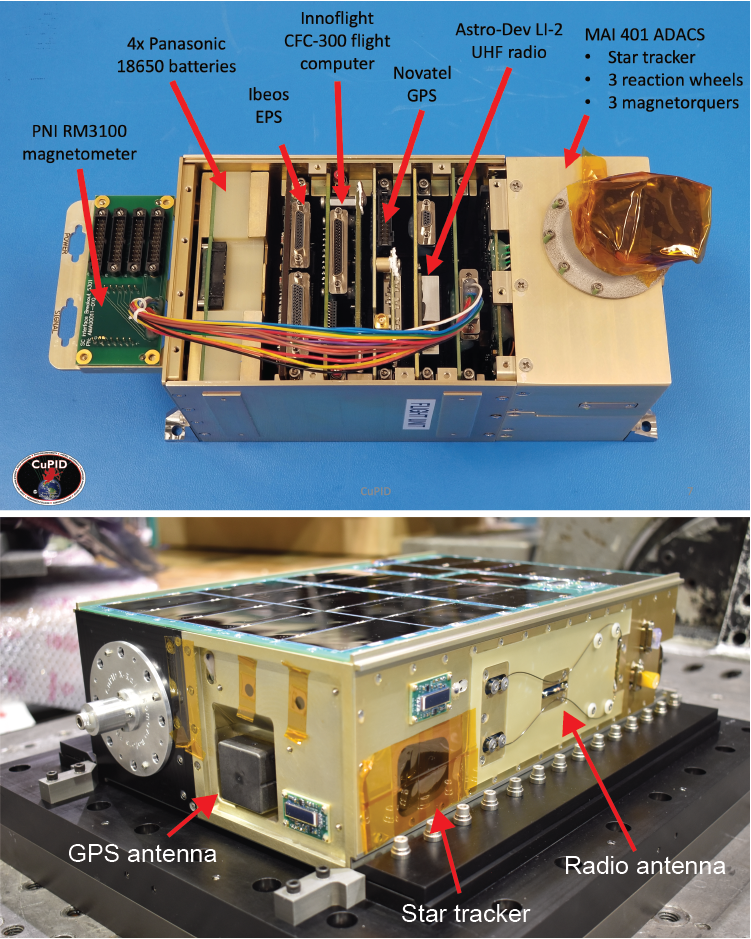}
\centering
\caption{\label{Avionics}Top: CuPID's avionics system and enclosure on workbench without wiring harness. The magnetometer is mounted on the backside of the diving board. The batteries supply the EPS which distributes power to the rest of the sub-systems. The flight computer receives data from the GPS and ADACS to control the spacecraft pointing. Ground commanding and telemetry is transmitted through the UHF radio. Bottom: CuPID mounted to a vibration table during environmental testing. Elements to note include CuPID's largest solar panel with 22 cells, the stowed radio antenna, and covered star tracker aperture leading to the avionics inside.}
\end{figure*}

\subsection*{Mission Critical Hardware and Software}
CuPID used an avionics system developed by Adcole Maryland Aerospace (AMA) that incorporated a number of components built by the company and some integrated from other vendors. The avionics enclosure and sub-systems are shown in Figure \ref{Avionics}. The avionics included Adcole Maryland Aerospace's MAI401 for attitude determination and control (ADACS - Attitude Determination and Control System). The ADACS was designed to determine spacecraft attitude with a magnetometer, six coarse sun sensors, and a star tracker, and provide pointing and stabilization with three reaction wheels and three magnetorquer rods.

The avionics used an Innoflight CFC-300 flight computer. CuPID's software, written in C++ with a modular design, was compiled for the Linux system running on the CFC-300. The modular flight code was divided into a number of modules controlling different elements of the spacecraft functionality. A \textit{Process Monitor} module was responsible for starting the satellite and the other software modules, and the \textit{Rules} module read current spacecraft data and managed issues when they arose. Some of the benefits and challenges of the \textit{Rules} module are discussed in a later section. The CFC-300 used NAND flash memory that was divided into three partitions. First was a non-volatile portion that contained the Linux operating system. The other two sections were regularly read from and written to, and contained the modular flight software in a smaller partition and generated spacecraft and science data in a larger partition. 

The communication system was built around matching Astronautical Development Lithium-2 (Astro-Dev LI-2) radios on the spacecraft and in the ground station communicating at 400.5 MHz. At 400.5 MHz, in the \textit{space research} ultra high frequency (UHF) band, CuPID was not allowed to send beacon packets. CuPID's antenna was a 1/4 wavelength deployable dipole antenna. The antenna was deployed by a burn-wire and resistor system that operated after ejection from the rocket.

\begin{figure*}
\includegraphics[width=400px]{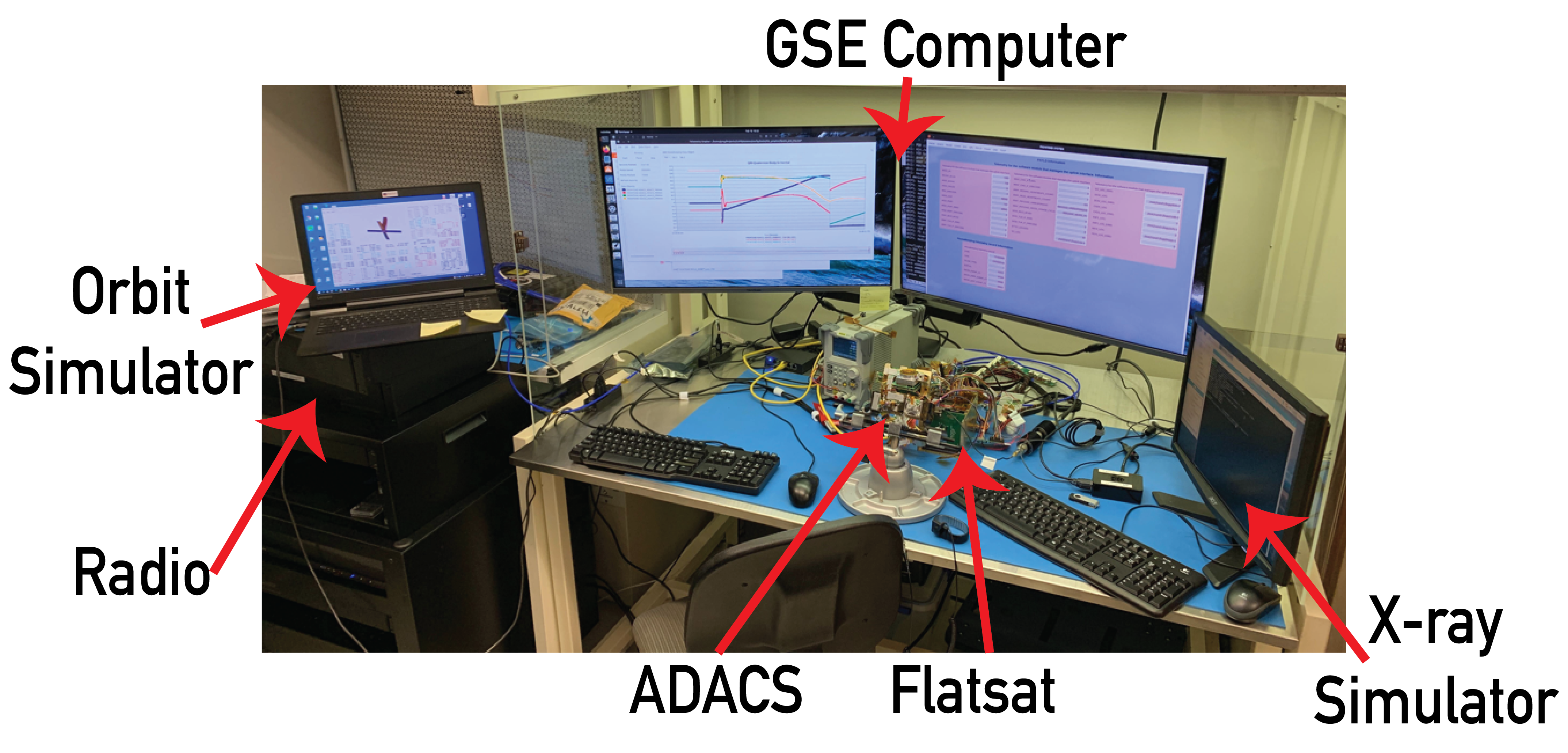}
\centering
\caption{\label{Flatsat}CuPID's flatsat used to test mission operations and software changes. An orbit simulation was input to the ADACS. The flatsat and ADACS responded accordingly. The operator monitored flight-like telemetry on the computer output and commanded the satellite through the radio.}
\end{figure*}

In addition to the development of a flight unit, a complete replica of the flight unit called the \textit{flat-sat}, was used for software and operations testing. See Figure \ref{Flatsat}. Additionally, the flat-sat could be used for debug purposes after launch of the flight unit.

\subsection*{Significant Mission Impacts}
The CuPID mission was significantly affected by the onset of the CoVID-19 pandemic. In December 2019, hardware elements of the flight unit were integrated at GSFC, and system level testing began. The CuPID team was notified the spacecraft was manifested on a launch scheduled for one year later, December 2020. When GSFC closed its gates in March 2020 due to the pandemic, CuPID was locked inside and still had a delivery date of September 2020, though later that month, the CuPID team was notified the launch was delayed until March 2021.

Through some political navigation, CuPID was extracted from GSFC and brought to BU where testing could more easily be conducted with special health and safety protocols. Testing facilities that were at GSFC had to be built at BU, and supply chain challenges and lead times forced the team to make compromises in certain testing hardware, such as an X-ray beam line.

During the summer of 2020, AMA was sold and CuPID's avionics support was terminated. The engineers working on CuPID moved to other companies. The BU CuPID team maintained a strong relationship with one of the lead engineers, and without this connection, CuPID would have had a significantly steeper climb to launch.

Testing of the integrated instruments and avionics were conducted daily. Flight software was iterated and tested on the flatsat, and instrument calibration was conducted on the flight unit. In November 2020, one month before planned delivery, another launch delay was announced, moving launch to September 2021. 

\section*{Anomaly Testing and Results}

The following sections outline some of the challenges the CuPID team navigated with the small satellite and investigated while attempting to contact CuPID. Each section leads with a synopsis of the content.

\subsection*{Early TLE's}
\textbf{Summary:} Multiple sources supplied multiple TLE's shortly after deployment which posed challenges in tracking multiple objects in the sky. As a secondary payload, information regarding the deployment and tracking of the satellite came with large error bars.\\
\textbf{Solution/Result:} Wider ground antenna beam width allows for more sky coverage, and given enough time, TLE's mature and are more precise.

Shortly after launch, when contact had not been made with the spacecraft after the first several ground-passes, the team began anomaly investigation. The first challenge the team encountered was identifying the proper TLE \footnote{Two Line Element} which corresponded to CuPID. Multiple sources provided different TLE's.

Within hours of launch and from the launch provider, the team was to get an estimated TLE from models based on the launch vehicle position. The delivery of this TLE was delayed one day by the launch provider. A second TLE was expected to be delivered from models run by GSFC. Another TLE was generated by the US Space Force North American Aerospace Defense Command (NORAD) which uses radar tracking of all objects orbiting Earth. In addition, not all objects ejected from the Centaur upper stage were claimed, therefore six possible positions of CuPID were identified. 

\begin{figure*}[ht]
\includegraphics[width=400px]{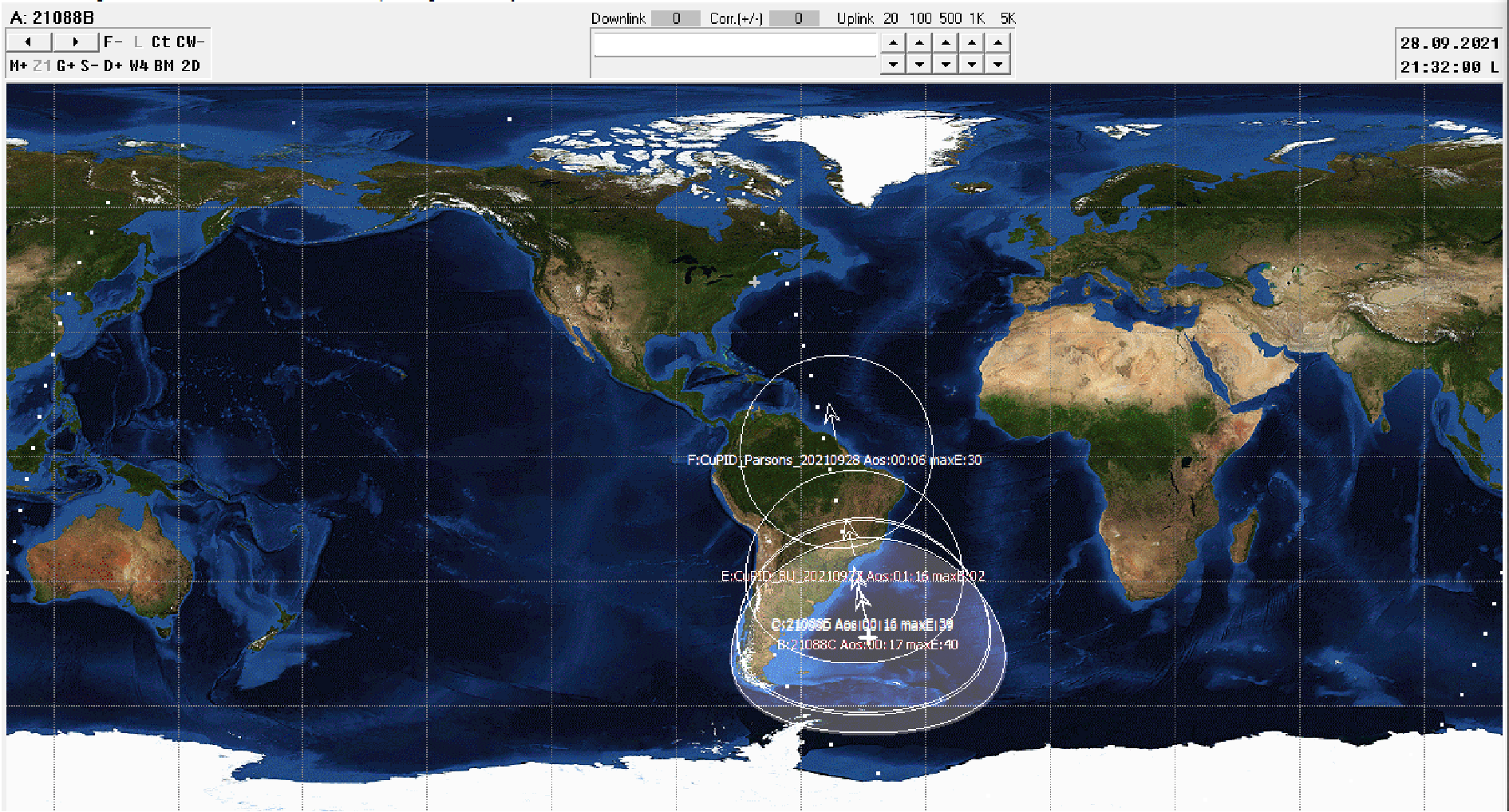}
\centering
\caption{\label{Possible}Computer ground track of potential TLE's for CuPID two days after launch.}
\end{figure*}

Figure \ref{Possible} shows the SatPC \footnote{SatPC was the software used to track TLEs and control the ground station antenna pointing. Link to software: https://www.amsat.org} map of the possible CuPID locations shortly after launch. Although all are relatively in track, the sequence of them proved challenging for ground pass operations because there were multiple satellites in the sky, at different locations.

A ground pass with multiple TLE's to track will have one satellite inside the RF beam, and multiple outside of it, leaving limited time to track each TLE. CuPID's ground station used two two yagi antennae to produce a beam width of about 30$^\circ$ and RF gain of 15 dBi. One antenna has a beam of about 60$^\circ$ and RF gain of 11.5 dBi. 

One antenna was removed to open the beam width, at the cost of RF power. This wider beam width allowed for greater inaccuracies in pointing and TLE knowledge. After a few days, the TLE's matured from accurate radar measurements. In addition, some of the objects were claimed by other institutions. CuPID's TLE was well known after a week but did not respond to commanding.

Should CuPID have had a beacon (discussed later) narrowing the TLE options would have been easier. With a satellite beacon, the ground station could listen to each TLE for a signal, and the maximum signal would designate the correct TLE.

\subsection*{B-dot control}
\textbf{Summary:} Magnetometer/magnetorquer control in the ADACS could have a mismatch in the hardware directions and controller causing excess spin rates of CuPID.\\
\textbf{Solution/Result:} Ground testing showed the directions were correct, though the polarity of the magnetorquers could not be confirmed, and testing of the complete B-dot control process showed inconclusive results. Ground imaging of CuPID on orbit were also inconclusive.

After deployment from the rocket, or after any power cycle or reset, the spacecraft ADACS was programmed to enter a rate nulling mode where it used a B-dot\footnote{\textit{B-dot} is a shorthand writing of the derivative of B, the rate of change of magnetic field magnitude} control to decrease the rate of spin of the spacecraft. It would remain in this mode until the spacecraft was commanded from the ground.

If, rather than nulling the spin rate, the controller spun the spacecraft to higher spin rates, it was anticipated the team would not be able to communicate with the spacecraft. The communication system required spin rates less than 10$^{\circ}$/s to successfully receive commands from the ground. Another cubesat mission using the same version of the ADACS hardware reported their spacecraft spinning up to 1200$^{\circ}$/s in one axis when in B-dot mode through monitoring beacon packets \cite{dolan2022operating}. 

One possible cause of such a failure is mislabeling of the polarity of the axes on the magnetometer or magnetorquers. Another cubesat group using the same MAI401 hardware did find some elements mislabeled in their system during ground testing. In the case of CuPID, the polarity of the magnetorquers was confirmed with ground testing of the flight unit prior to launch and on the flat-sat after launch. A second possible cause of this failure could be issues in the firmware that performs the B-dot algorithm on the ADACS unit. The algorithm requires input from magnetometer measurements and results in output of magnetorquer actions. A coordinate system issue in the firmware could not be tested with the available tools and still remains a possibility. The B-dot control of the flight unit CuPID was tested on an air bearing setup, though results were inconclusive because of the total mass and friction of the air bearing assembly. 

Some resources exist to image orbiting bodies which can provide information on the behavior of a spacecraft, such as spin rate or cross section, even if the spacecraft is not responding. Optical imaging was explored to monitor CuPID's spin rate, but the size of the 6U spacecraft, with no deployable solar panels to expand the cross-section, produced a small signal to noise, and ultimately spin rates were not reported back to the CuPID team.

\subsection*{Memory corruption}
\textbf{Summary:} CFC-300 data partition became corrupted twice on the flight unit spacecraft due to a power down during write moment on drive.\\
\textbf{Solution/Result:} Power down sequence would un-mount partition before powering down the system.

Over the course of months of power cycling and daily testing, the partition for data storage, but not flight software became corrupt two times. Through direct commanding, of the CFC-300 computer, the corrupt data partition was erased and reset. Flight software was not disrupted. Additional abilities were developed to allow the flight software and a ground operator command the CFC-300 in case this happened on orbit.

The cause of the corruption was determined to be an electrical power anomaly during a moment of data writing. The power anomaly was found to be caused during a operator defined power down of the spacecraft at an unfortunate, but precise moment. The slow power drop during a write action would cause the corruption. This problem was remedied with a software defined un-mounting of the partition of the hard drive before powering down.

Although the problem was corrected during manual shut down sequences, unexpected resets could still cause this problem. On orbit, unplanned power events can be common, and could cause the hard drive to become corrupt. Although the flight software partition never corrupted during ground testing, it is still a possibility on orbit.

\subsection*{Spacecraft reset timing}
\textbf{Summary:} Rules in the software could reset the spacecraft in a periodic fashion that would synchronize with ground passes.\\
\textbf{Solution/Result:} Ground passes were conducted from Fairbanks, AK, with a mobile ground station to change the location, and increase the number of passes in a day.

The \textit{Rules} module included a few software defined rules to control the spacecraft and maintain it's state of health.
\begin{enumerate}
    \item Every 12 hours without ground commands- Entire spacecraft reboot
    \item Every 6 hours- radio reset
    \item Battery below 60\% state of charge - shut down everything except the power monitoring system until power returns
\end{enumerate}

Should the satellite receive no successful commands from the ground in 12 hours, the flight computer would force the electrical power system to reset, which then power cycled the flight computer and restarted the entire satellite. Spacecraft startup took about two minutes. The 12 hour reboot was included as a precaution in case an issue on orbit required a reboot to maintain radio link.

\begin{figure*}[ht]
\includegraphics[width=270px]{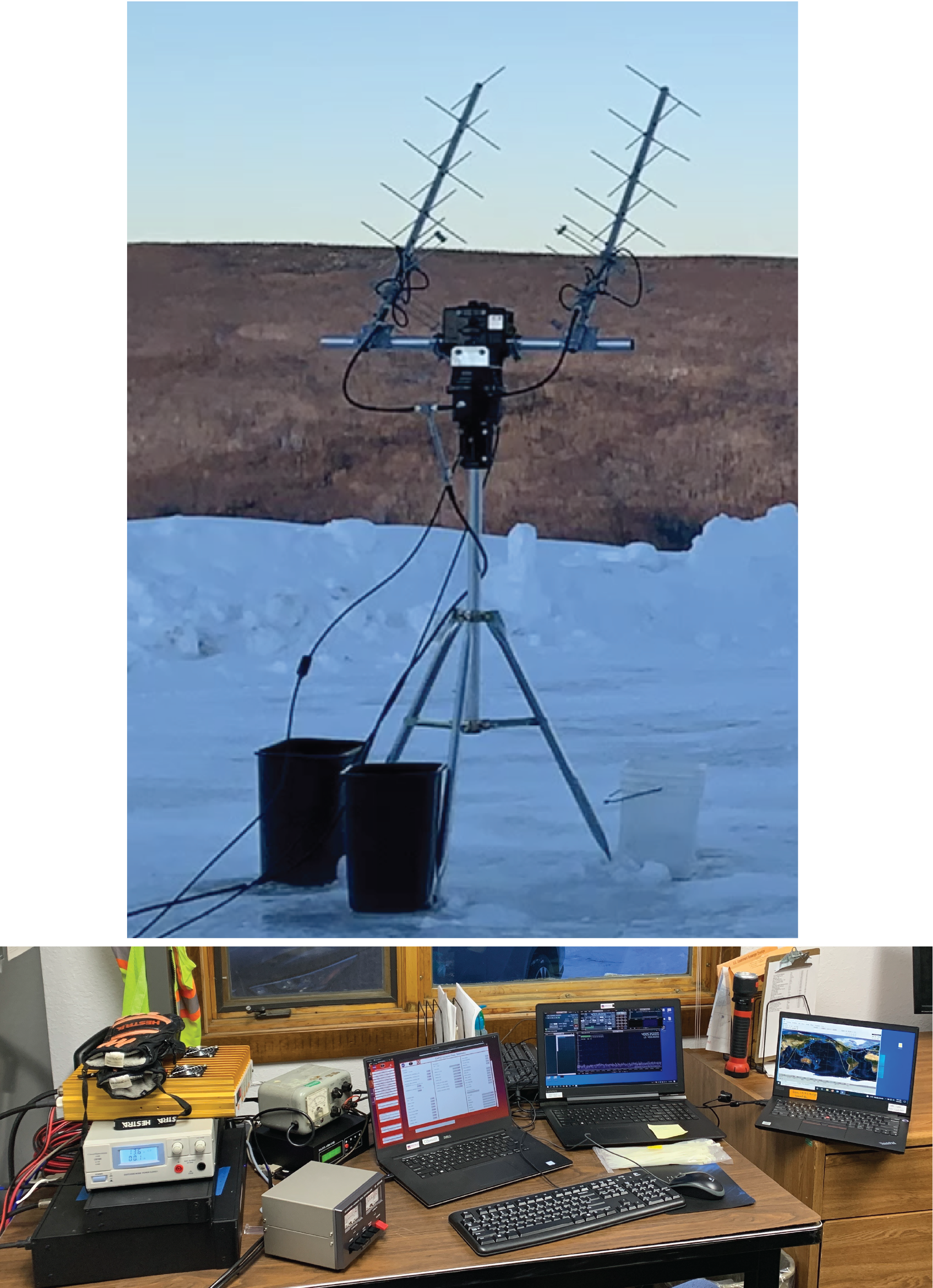}
\centering
\caption{\label{Mobile}The mobile ground station during testing at Poker Flat in Jan. 2022. TOP: Tripod, rotator head and antennae. BOTTOM: Radio, RF amplifier, power supplies and laptop setup inside.}
\end{figure*}

The 6 hour radio reset was included after ground testing proved the radio required frequent resets. The radio showed instances of poor packet management, and at times was locked transmitting noise. In these instances, no commands could enter the radio via RF, and the radio was consuming battery power while transmitting. The software defined radio reboot would cut these issues short should they occur. The reset only lasted 10 seconds.

A rule regarding the battery state of charge was included to protect the battery health. CuPID utilized two body mounted solar panels, one on each of the 6U faces of the satellite. Software would power off the satellite below a 60\% state of charge, and would power it back on as the charge increased. It was determined that in a tumbling attitude and only using the limited power consumption after startup, CuPID was likely power negative. Modeling showed this power cycling could occur once per day.

Although the concept of these rules are common on satellites, the scheduling provided an unfortunate mismatch for CuPID's orbit. CuPID's orbit at 550 km has a period of almost exactly 90 minutes, and because it is sun-synchronous, ground passes were at nearly the same time each day. Each day of ground passes consisted of four total ground passes, with two iterations of two sequential orbits (e.g. 10AM, 11:30AM, and 10PM, 11:30PM). There \textit{could} be a condition where the satellite would be powered off during many sequential ground passes because the 90 minute orbit period is a factor of the 6 hour radio reset and 12 hour spacecraft reboot. This could have been prevented by selecting a reset interval that was not an interval of the orbital period.

In order to decrease the likelihood that this was the cause of no-contact, multiple sequential orbit ground passes needed to be conducted. At higher latitudes, the 97.6$^\circ$ orbit inclination allowed for up to 8 sequential orbit passes, although some were at elevations below 10$^\circ$. The team developed a mobile ground station with a small antenna rotator head, tripod and some power supplies, see Figure \ref{Mobile}. The rotator was a Yaesu G5500 controlled by an EA4TX ARS-USB rotator controller. This mobile station was brought to Poker Flat Research Range outside of Fairbanks, AK in January, 2022 for six days of testing. Seven consecutive passes were conducted on two out of the six days of operation from Poker Flat Research Range. Although multiple sequential passes were conducted, CuPID did not respond to commanding.

\subsection*{Antenna deployment}
\textbf{Summary:} The antenna was designed to be deployed with a burn wire mechanism, and if not deployed, the antenna gain was limited to 1.\\
\textbf{Solution/Result:} Ground testing showed confidence in the mechanism, though the issue on orbit is still a possibility and cannot be confirmed.

The deployment of the antenna is an unlikely cause of no signal from CuPID. The deployment mechanism was successfully tested after a vibration environment four times (risk reduction of the chassis, low level vibe, requirement vibe, increased requirement vibe) and performed nominally each time. The stowed antenna can be seen in Figure \ref{Avionics} and Figure \ref{RadioStuff}A. The deployment software and mechanism from boot-up was tested multiple times without failure. 

A less significant but possible software issue regarding the antenna deployment, was that the software would only attempt to deploy the antenna once at initial start up after deployment. After the automatic software attempt was made to burn the wire, a file was written that told the software the antenna was deployed. Any later startup, would then read this file and know not to deploy the antenna. There was no automatic method to read a failed antenna deploy and re-attempt. Link budgets showed that contact could still be made, though margins were closer because the deployed antenna gain was 2.15 dBi verses stowed gain at 1 dBi. Ground testing (although higher power than on orbit) proved a stowed antenna would still communicate. Commands from the ground could attempt to re-deploy the antenna. A better design of the antenna deployment system would include automatic repeated attempts to deploy the antenna should the satellite sense it failed.

\subsection*{Radio performance}
\textbf{Summary:} The LI-2 radio likely had a mismatch in firmware and hardware components, which manifested in poor performance.\\
\textbf{Solution/Result:} Ground testing illuminated the poor performance, though the cause was not determined until days before delivery. Further ground testing could have dis-qualified the radio for flight.

\begin{figure*}[ht]
\includegraphics[width=350px]{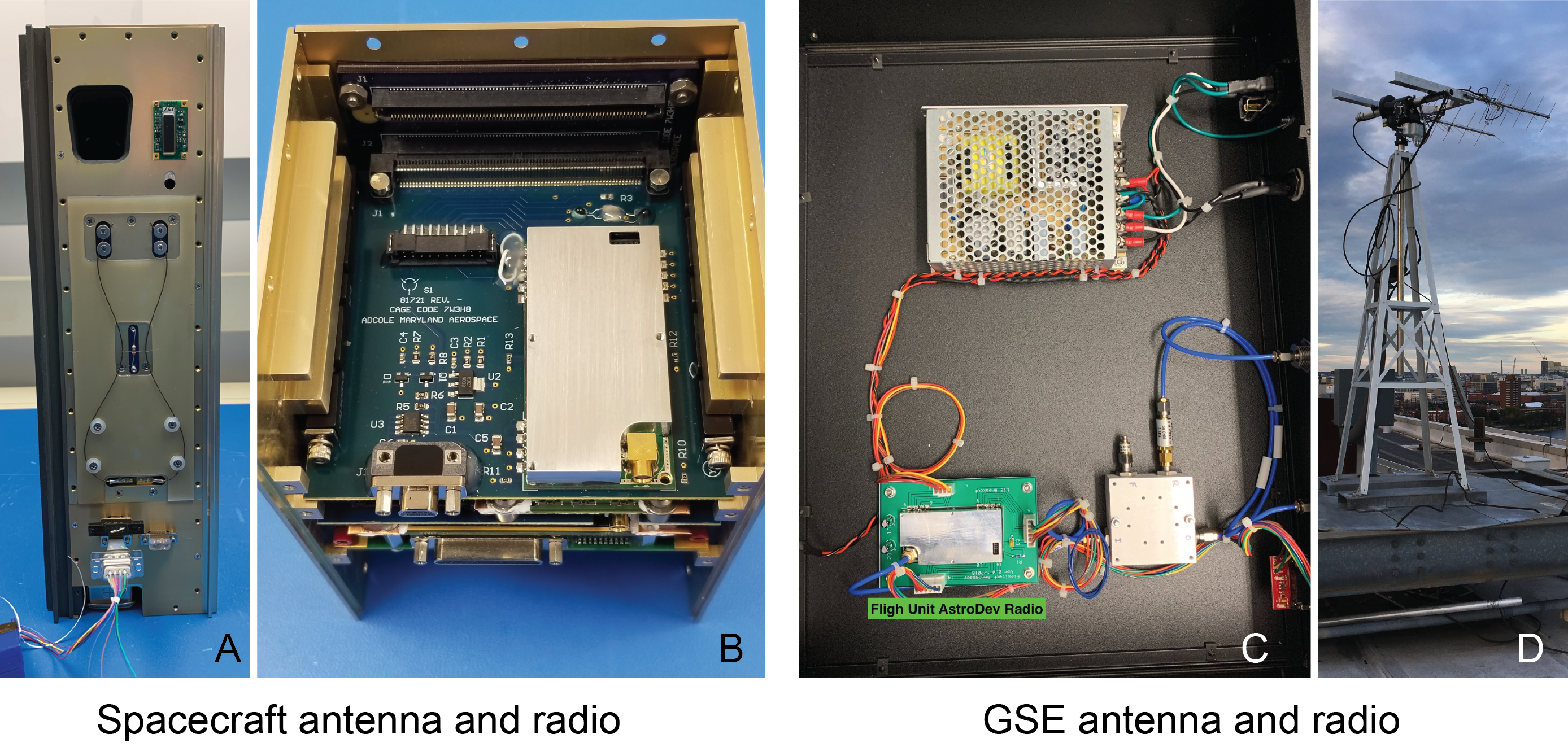}
\centering
\caption{\label{RadioStuff}CuPID's spacecraft radio (left) and ground radio (right) systems. A: CuPID with stowed antenna. B: CuPID's LI-2 radio and board inside avionics. C: Ground radio system. D: Ground antenna.}
\end{figure*}

CuPID's communication system was based around a matched pair of Astro-Dev LI-2 radios. One radio was configured for the CubeSat, the other for the ground station. Licensed to communicate at 400.5 MHz, the ground radio transmitted at 9,600 bits/s and received at 38.4 kbits/s. The spacecraft transmitted and received the opposite baud rates. The spacecraft's 1/4 wavelength deployable antenna (Figure \ref{RadioStuff}A) was deployed by a burn wire resistor. The spacecraft flight computer was configured to communicate with the radio at 115,200 baud.

Discussions with the radio manufacturer days before the delivery of CuPID illuminated a possible cause of failure. The radio's default interface baud rate was 9,600, whereas the flight computer was configured to communicate with the radio at 115,200. On the ground, the radio was configured, and stored to communicate at 115,200 baud. Should the radio reset to default values (for example, from a single event upset), the flight computer had no means to determine and reset the baud rate to re-establish the interface.

Testing of the radio on the ground consistently showed an offset to the transmit frequency. Although configured at 400.5 MHz, the radio transmitted at 400.531 MHz. This was originally presented to the radio manufacturer in March of 2020, but was not answered until the discussion days before delivery of CuPID. The frequency offset was described as a signature of a mismatch between the firmware and hardware on the radio due to a version control setback from supply chain issues during manufacturing.

Licensed in the space research band, CuPID was not allowed to beacon data. Instead, CuPID's communication system required a command to enter the spacecraft before data was transmitted from it. This means that a significant signal must reach the spacecraft radio to be decoded and acted upon. The input signal must be above the noise floor of the radio. An increase in the ground station power output would increase the power input to the spacecraft radio. An RF amplifier was added into the ground station (Toptek PA-100-05) that output up to 100 W to the ground station antenna (antenna gain of about 15 dBi). Assuming a loss of about 5 dB in transmission, 100 W from the amplifier resulted in an effective isotropic radiated power (EIRP) of 60 dBm, or 1000 W. This emitted power was 20 times the designed and expected link budget. This still did not result in response from CuPID. The power amplifier was also used in passes at Fairbanks, AK. 

The team also performed ground passes from NASA Wallops Flight Facility (WFF) with a 30 m dish antenna. The antenna has a gain of 36 dBi. CuPID's ground radio input 1.75 W, resulting in an EIRP of 63.43 dBm or 2203 W (assuming 5 dBm of losses). Although a sizeable emission, this still did not produce a response from CuPID. Seven ground passes over 4 days were conducted from the WFF ground station.

Discussions with the IT-SPINS CubeSat team from Montana State University (MSU) showed they too had problems with an Astro-Dev LI-2 radio procured near the same time as CuPID. Not only did IT-SPINS B-dot control in the MAI-401 ADACS accelerate the spin rate of the satellite, but their LI-2 would not take commands, although they did have a radio beacon transmitting state of health data \cite{dolan2022operating}.

CuPID had spare radios that were purchased during development. Two radios were sent to MSU for characterization. The noise floor of the radios was found to be unstable, varying between -130 dB and -90 dB when tested in an anechoic chamber with a known noise floor at -123 dB. This could be an artifact of the unproven mismatch between hardware and firmware in the radio. A noise floor of -90 dB is significantly high. Should this be real, the Gain/Noise factor of the radio would be at least 40 dB lower than the designed value.

In terms of link budget, one can make estimates from a best case scenario ground pass. For a high elevation pass ($>$70$^\circ$) where the free space path loss was a minimum, and from WFF where the EIRP was maximized, link budget estimates with the as measured radio performance suggest that uplink margins were not favorable. The power into the radio would be near -104 dB, which is lower than a noise floor at -90 dB. Estimates for the bit error rate (BER) calculated from the noise temperature and received figure of merit show that link could be made, but margin was within a few dB. 

Three things that can further decrease the BER during a pass include spacecraft orientation, elevation and occupied bandwidth.

There are a few spacecraft orientations which decrease CuPID's antenna gain. The ideal gain was 2.15 dBi. If CuPID was spinning, it is likely this value was decreased, therefore decreasing the BER margin. Additionally, if CuPID's antenna never deployed, the receiver gain was reduced to 1.

At 70$^\circ$ elevation, the free space path loss is -139.75 dBm, whereas at 10$^\circ$ elevation, the free space path loss is -151 dBm. This decreases the BER margin by 12 dB.

Occupied bandwidth was measured at -3 dB from center frequency peak. The frequency offset of the radio (about 30 kHz) would require a wider occupied bandwidth. CuPID's FCC license reports the bandwidth at -40 dB and -60 dB, at 22.9 kHz and 216.9 kHz respectively. At a -40 dB bandwidth, the occupied bandwidth was 43.44 dB. This is 10.43 dB larger than the bandwidth at 3 dB, and would decrease the BER margin accordingly.

These situations all decrease the BER margin, and all are likely included in some capacity. Even with the strongest signal and best BER margin, from WFF, the margin would likely be consumed by these additional taxes on the uplink system.

\section*{Improvements}

The development and operations of CuPID led to multiple lessons learned that are useful to future missions. Here, three improvements to the CuPID mission are discussed.

First, a radio should be tested beyond what would be expected on orbit. CuPID's radio was assumed functional from the manufacturer, and issues during development were attributed to the development team's software. In reality, the radio performance was far below what was expected. Although the radio had functionality on the ground, issues such as the low data throughput, and frequency offset should raise flags to test the radio further. The noise floor of the radio should be confirmed before launch. In addition, the default values of the radio should be coded into a software check for the flight computer to try should commands not be received.

Beaconing is a communication feature that CuPID's radio license was not permitted to use. Beaconing, is a radio frequency (RF) communication method where a satellite emits packets of information at known period. The Federal Communication Commission (FCC) does not allow beaconing outside of the \textit{amateur} frequency band (UHF 420.0 to 450.0 MHz). CuPID, at 400.5 MHz was not allowed to beacon. For a mostly student built satellite, a radio beacon should be an essential component.

A great benefit of the beacon, is that it allows the amateur radio community to listen to the satellite, but not de-code any data unless the process is shared with them. The SatNOGS Network\footnote{https://network.satnogs.org/}, is a community of amateur radio users who use the network to track satellites and upload their ground pass observations. The network has ground stations all over the world, and data can be accessed online. Combining the SatNOGS Network with a beaconing satellite would have allowed CuPID to get nearly continuous monitoring. Additionally, a beacon would allow a mission in early operations to quickly locate which TLE corresponds to a team's satellite. By instead listening to each TLE for the beacon, one could determine their TLE out of multiple options based on signal strength.

A third improvement to CuPID would have included more automatic rules in flight software that execute if the satellite does not receive commands from the ground. Although a simple reset could fix some issues, commands that attempt to re-deploy the antenna, or re-establish communication with the radio could have been implemented.

\section*{Current Cubesat Economy}
The small satellite industry continues to rapidly evolve and grow, particularly in the commercial and government markets. Launches of small satellites have increased near exponentially \cite{villela2019towards}. A feature of this growth is rapid commercial realignments including mergers and acquisitions. In some cases this can have significant implications for customers and be disruptive on schedule and program development. In the case of CuPID, the avionics developer was acquired and the branch responsible for the avionics development dissolved prior to the end of the program, causing a rapid re-alignment of development activities within the CuPID team. Stability should be an element of consideration in addition to performance and cost when evaluating vendors.

\section*{Conclusion}

On-orbit anomaly testing of the CuPID CubeSat led to a few important lessons learned. These are some critical elements for developing a CubeSat:
\begin{enumerate}
    \item Having a flat-sat, or replica of the spacecraft electronics that a builder interacts with like it is the flight unit is indispensable for ground testing.
    \item Test a flight radio intensely - confirm noise floor, etc.
    \item Put licensed radio frequency in a frequency band that allows beaconing. Early operations are easier if a satellite automatically transmits.
    \item If using NAND memory, include software to un-mount the memory before powering off the satellite to prevent corruption of the drive partition.
    \item Include and test software fail-safes that are acted upon in the case of no-contact from the ground. (E.g. automatic re-attempt of antenna deploy, reset of radio baud rate)
\end{enumerate}

The CuPID cubesat was built with the objectives to develop a small form-factor and low resource wide field-of-view X-ray telescope and to use this instrument to understand macro-scale spatial and temporal properties of magnetopause reconnection. Ground characterization allowed demonstration of the performance of the miniaturized X-ray telescope. After the spacecraft was launched into orbit, an anomaly in-flight prevented communication from the ground-station with the spacecraft. Further anomaly testing spanned a number of spacecraft sub-systems and found irregular and temporally variable performance of the radio as the most likely cause of failed communication.  
\newpage

% \begin{table}[h] % [h] forces the table to be output where it is defined in the code (it suppresses floating)
% 	\caption{Example table.}
% 	\centering
% 	\begin{tabular}{l l r}
% 		\toprule
% 		\multicolumn{2}{c}{Name} \\
% 		\cmidrule(r){1-2}
% 		First Name & Last Name & Grade \\
% 		\midrule
% 		John & Doe & $7.5$ \\
% 		Richard & Miles & $5$ \\
% 		\bottomrule
% 	\end{tabular}
% \end{table}

%----------------------------------------------------------------------------------------
%	BIBLIOGRAPHY
%----------------------------------------------------------------------------------------

\bibliographystyle{unsrt}

\bibliography{references.bib}

%----------------------------------------------------------------------------------------

\end{document}